\documentclass[gmd, manuscript]{copernicus}

\usepackage{algorithmic}
\usepackage{algorithm}

\begin{document}

\title{Unified sparse framework for large-scale simulations using the material point method}


\Author[1,2,3]{Yidong}{Zhao} 
\Author[4]{Lars}{Blatny}
\Author[5]{Xiang}{Feng}
\Author[1,2,3]{Mikkel M.}{Juel} 
\Author[5]{Chenfanfu}{Jiang} 
\Author[1,2,3][jgaume@ethz.ch]{Johan}{Gaume}

\affil[1]{Institute for Geotechnical Engineering, Department of Civil, Environmental and Geomatic Engineering, ETH Zurich, Zurich, 8093, Switzerland}
\affil[2]{WSL, Institute for Snow and Avalanche Research SLF, Davos Dorf, 7260, Switzerland}
\affil[3]{Climate Change, Extremes, and Natural Hazards in Alpine Regions Research Center CERC, Davos Dorf, 7260, Switzerland}
\affil[4]{Sorbonne Universit\'e, CNRS - UMR 7190, Institut Jean Le Rond d'Alembert, Paris, France}
\affil[5]{Artificial Intelligence and Visual Computing Laboratory, University of California, Los Angeles, United States}




\runningtitle{TEXT}

\runningauthor{TEXT}

\received{}
\pubdiscuss{} 
\revised{}
\accepted{}
\published{}


\firstpage{1}

\maketitle

\begin{abstract}
The material point method (MPM) is a hybrid particle-grid method widely used for large deformation problems with history-dependent behavior, including geophysical mass flows.
Standard MPM often relies on a dense background grid, which can be highly inefficient when material occupies a small fraction of the computational domain.
Such sparsity is common in many large-scale geophysical mass flow problems.
Here, we introduce a unified sparse background-grid framework for large-scale MPM simulation. 
The framework treats sparse grid construction as a general active-node indexing problem.
We develop two architecture-specific implementations to realize the same sparse framework: a scan-based strategy for CPUs and a hash-based strategy for GPUs. 
Through benchmark problems and a large-scale landslide simulation, we show that the framework provides identical results as standard dense MPM while reducing computational time and memory usage by one to two orders of magnitude in strongly sparse cases.
\end{abstract}


\introduction  
Geophysical mass flows are gravity-driven movements of soil, rock, snow, ice, water, or their mixtures.
Landslides, rock avalanches, debris flows, and snow avalanches are typical examples of geophysical mass flows.
These flows can travel over long distances and cause severe damage to infrastructure and communities~\citep{iverson1997physics,schweizer2003snow,hungr2014varnes,lacroix2020life}. 
Numerical simulation is therefore an important tool for understanding their dynamics and supporting hazard assessment. 
However, efficient simulation of such flows is challenging because they involve large deformation, rapid motion, and complex interactions with large terrain domains.

The material point method (MPM), typically attributed to~\citet{sulsky1994particle}, is a hybrid particle-grid method well suited for simulating history-dependent materials undergoing extreme deformation.
In MPM, the material is discretized using Lagrangian material points (particles), while the balance equations are solved on an auxiliary Eulerian background grid.
This combination avoids severe mesh distortion under large deformations while enabling the ability to readily track material history. 
Owing to these advantages, MPM has been applied to a broad range of geophysical-flow problems, including granular flows, landslides, debris flows, and snow avalanches~\citep{fern2019material,GAUMEdynamic2018, GAUMEinvestigating2019, krzyzanowski2021modelling, CICOIRApredictive2022, KOHLERmpm2022, KYBURZpotential2024, ROUSSEAUtransition2024, BLATNYobservations2026}.
Several open-source implementations have also been developed in languages such as C\texttt{++}, MATLAB, Python, and Julia~\citep{germain2000uintah,nairn_mpm,zhang2016material,kumar2019scalable,hu2019taichi,wyser2021explicit,de2021karamelo,anura3d,shi2024geotaichi,huo2025high,blatny2025matter,zhao2026geowarp}. 
They have made MPM easily accessible to the broad geomechanics and geophysics communities.

Despite its popularity, the background grid, which is the key ingredient of MPM may become a major source of computational inefficiency. 
In standard implementations, the grid is initialized as a bounding box, covering the potential region that particles may occupy during the whole simulation steps.
For many large-deformation mass flow problems, this potential region can be very large as the material may travel far and spread widely from its initial position.
At any given time, however, the moving material typically occupies only a small fraction of the full domain (\textit{i.e.}, the occupied domain is sparse).  
As a result, many grid nodes are inactive. 
They receive no particle contribution and do not participate in the physical update. 
Allocating and looping over these inactive nodes leads to unnecessary memory usage and computational cost. 
This mismatch becomes extremely severe when the potential domain is large but the material distribution remains localized or spatially sparse.

Several studies have recognized this spatial sparsity in MPM and have explored ways to reduce the cost of the background grid. 
A dynamic meshing technique~\citep{shin2009numerical} is introduced, in which grid quantities are allocated according to the particle distribution at every step. 
Those grid quantities are searched and stored through map-like data structures. 
However, the implementation relied on C\texttt{++} map containers and was not designed for parallel execution. 
Other developments have explored adaptive or multi-resolution MPM to improve efficiency through local refinement~\citep{tan2002hierarchical,ma2006structured,lian2015mesh,gao2017adaptive,cheon2019adaptive,he2025multi}. 
These methods are useful, but their main objective is not a sparse representation of active nodes. 
They also usually require significant changes to the MPM formulation such as interpolation functions, particle-grid transfer schemes, and grid data management. These modifications significantly increase implementation complexity.
In addition, spatial hashing and hierarchical grids have enabled efficient storage and computation~\citep{gao2018gpu,hu2019taichi,wang2020massively,qiu2023sparse,chen2025sparse}. 
These works demonstrate the potential of exploiting sparsity, but they are built around either GPU data structures or specific libraries such as Kokkos and Cabana~\citep{edwards2014kokkos,trott2021kokkos,mniszewski2021enabling,slattery2022cabana}.

These existing works show the necessity for a sparse framework that is not tied to a particular data structure or hardware platform.
Such a framework should leave the underlying MPM formulation unchanged (that is, the particle-grid transfers and the solution of the governing equations). 
Only the storage and access of grid data need to be modified.
At each time step, the nodes influenced by particles form an active node set.
By assigning each active node a compact index, nodal quantities can be stored and updated only over this reduced set.
From this viewpoint, sparse MPM becomes an indexing problem: the structured background grid indicates the physical node locations, while a compact indexing map defines where the corresponding nodal data are stored in memory. 
This abstraction makes it possible to realize the same sparse framework using different (and more proper) algorithms on different hardware.

In this work, we propose a unified sparse background-grid framework for MPM. 
Instead of starting from a specific data structure, we formulate sparse MPM using two objects: the active-node set and a compact indexing map. 
The active-node set identifies the grid nodes influenced by particles, and the compact indexing map assigns these nodes storage locations.
The framework restricts memory allocation and grid-based computation to active nodes only, without changing the underlying MPM formulation. 
Because the framework is formulated independently of a particular data structure, it can be realized in different ways depending on the hardware platform.
We develop two implementations. 
The scan-based implementation is designed for CPUs. 
It constructs a binary activity mask over the candidate grid and applies a global scan to generate compact indices. 
The hash-based implementation is designed for GPUs. 
It avoids a global scan and instead constructs the active-node set through parallel insertion and lookup in a hash table.

This work makes three main contributions. 
First, it introduces a general sparse MPM framework by formulating sparse grid construction as a general active-node indexing problem.
Second, it shows how this framework can be realized efficiently on different architectures through scan-based and hash-based implementations.
Third, the implementations are released as open source to facilitate adoption in other MPM codes. Through examples from benchmark problems to large-scale landslide simulations, we show that the proposed framework reduces both computational time and memory usage by one to two orders of magnitude in strongly sparse cases. 
These results demonstrate that the sparse framework provides an effective route toward scalable simulation of a wide range of geophysical mass flow problems.
The remainder of the paper is organized as follows. We first give an overview of the proposed framework and show its performance using examples with different levels of sparsity. Then, we provide method details and compare the two implementations of the same underlying sparse framework.

\section{Results}
\subsection{Method overview}
\subsubsection{Material point method}
\begin{figure}[!ht]
    \centering
    \includegraphics[width=0.7\textwidth]{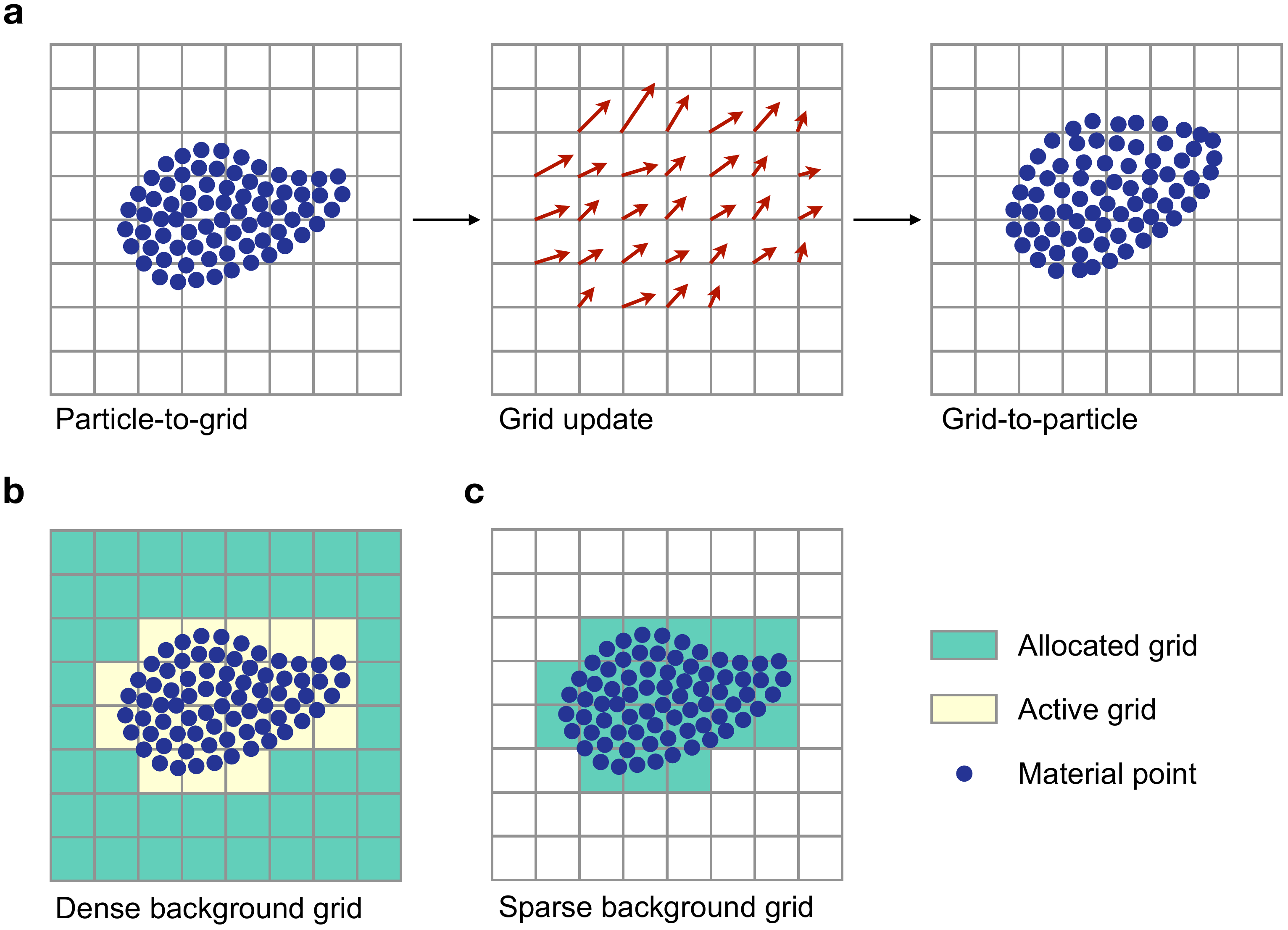}
    \caption{Material point method (MPM) overview in two dimension. a. Basic computational steps of MPM. Blue dots denote material points, squares represent the background grid used for computation, and red arrows indicate updated grid velocities. b. Dense background grid. The green region denotes the allocated grid covering the whole domain, while the yellow region highlights the active grid nodes at the current time step. c. Sparse background grid, where grid allocation is restricted to the active region, so that allocated and active grid nodes coincide.}
    \label{fig:mpm_overview}
\end{figure}

In MPM, the material is discretized using Lagrangian particles that carry history-dependent quantities (\textit{e.g.}, mass, velocity, and strain), while the governing equations are solved on an auxiliary Eulerian background grid.
As illustrated in Fig.~\ref{fig:mpm_overview}a, each time step consists of three procedures. 
First, particle quantities are transferred to grid nodes (particle-to-grid). 
The governing equations are then solved on the grid, where nodal quantities are updated (grid update). 
Finally, the updated grid quantities are interpolated back to the particles to advance their states (grid-to-particle).

During the particle-to-grid and grid-to-particle procedures, each particle interacts with the grid through shape functions defined on each node with local support. 
As a result, a particle influences only a small set of neighboring grid nodes. 
We refer to a grid node as active at a given time step if it lies within the support of at least one particle. 
The remaining nodes, outside all particle supports, are inactive and do not take part in the particle-grid transfers and physical update at that time step.

Because particles move through the domain, the set of active grid nodes changes over time. 
In standard MPM implementations, a sufficiently large background grid is constructed to cover the whole region of possible particle motion. 
However, as shown in Fig.~\ref{fig:mpm_overview}b, the region occupied by particles at a given time step often represents only a small fraction of the large grid domain. 
As a result, many nodes remain unused, leading to unnecessary memory usage for allocating and resetting nodal quantities, and additional computational costs when looping over nodes (\textit{e.g.}, when normalizing nodal quantities after particle-to-grid transfer).
This observation motivates a key question addressed in the following section: how can we construct a background grid that includes only the active nodes, as illustrated in Fig.~\ref{fig:mpm_overview}c?

\subsubsection{Sparse MPM}
\label{sec:sparse_mpm}
\begin{figure}[!ht]
    \centering
    \includegraphics[width=0.7\textwidth]{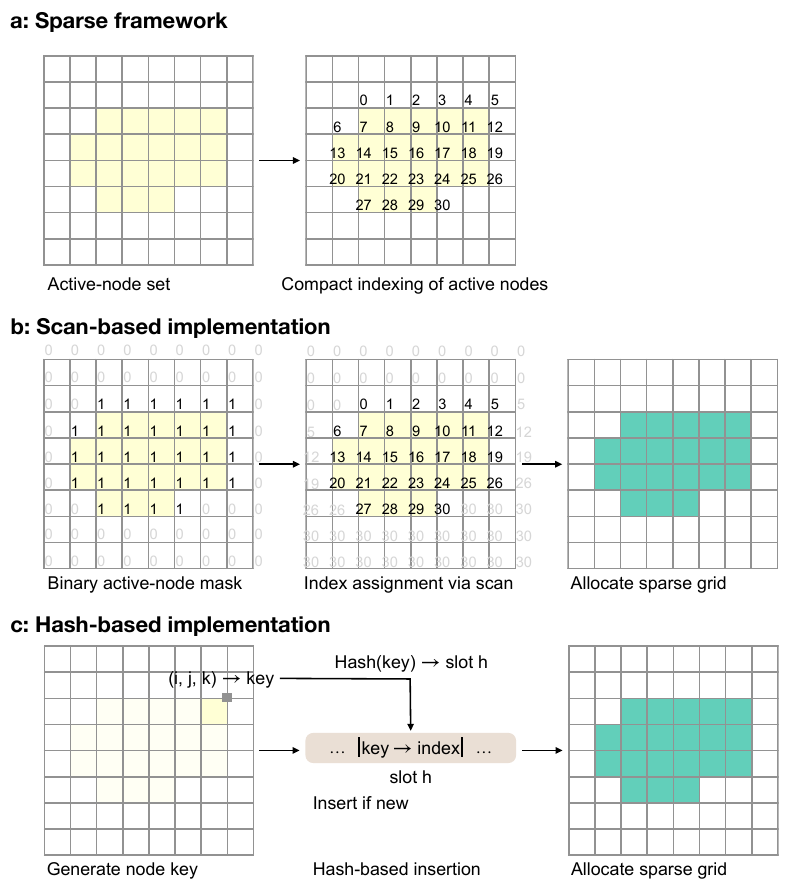}
    \caption{Unified sparse background-grid framework and its architecture-specific implementations. a. Sparse framework. Particles influence grid nodes within their support, defining a set of active nodes. These nodes are then assigned compact indices. These indices are used to allocate a sparse background grid that keeps only the active region. b. Scan-based implementation. A global binary mask is first constructed by marking all active grid nodes with value 1. Then a prefix scan is applied to assign compact and consecutive indices. With the indices, memory is allocated for only the active region. c. Hash-based implementation. A unique key is generated based on its grid location. Then the key is inserted into a hash table. If a key is encountered for the first time, the index is increased and assigned to that node. Once all active nodes are checked, the hash table contains the complete set of active nodes with their indices. Lastly, the memory is allocated only for the active region.}
    \label{fig:mpm_sparse}
\end{figure}

As shown in Fig.~\ref{fig:mpm_sparse}a, the sparse framework is built in two steps.
First, we identify the active-node set, namely the set of grid nodes influenced by particles at the current time step.
Second, we assign each active node a compact index from $0$ to $n_{\mathrm{active}} - 1$, where $n_{\mathrm{active}}$ is the total number of active nodes.
This compact index specifies where the nodal data (such as mass, momentum, velocity, and force) are stored in memory.
It maps each node from its physical position to a corresponding array location.
In this way, the grid node can still be identified by its physical location in the structured background grid, but its data are stored in a compact sparse array.
The MPM formulation itself is unchanged.
Particle-to-grid transfer, grid update, and grid-to-particle transfer are performed as usual, except that nodal quantities are stored and accessed through the compact index.

The key task is to construct the active-node set and its associated compact indexing map efficiently. 
This task can be achieved using different algorithms.
In this work, we develop two implementations for CPU and GPU architectures: a scan-based implementation (Fig.~\ref{fig:mpm_sparse}b) and a hash-based implementation (Fig.~\ref{fig:mpm_sparse}c). These are two realizations of the same framework rather than two separate frameworks.

The scan-based implementation (Fig.~\ref{fig:mpm_sparse}b) adopts a structured, grid-wide view of the problem. 
A binary mask is first constructed over the candidate grid by marking nodes influenced by particles. 
In this mask, a value of 0 indicates an inactive node, while 1 indicates that the node is influenced by at least one particle. 
Then, a prefix scan is applied to the mask to assign compact and consecutive indices (\textit{i.e.}, from 0 to $n_{\text{active}} - 1$) to all active nodes. 
Specifically, each active node receives a unique index based on the cumulative number of active nodes preceding it, while inactive nodes are skipped. 
With the indices, we can allocate memory only for the active region. 
Because the approach relies on a global scan operation to generate consecutive indices, we refer to it as the scan-based approach. 
This implementation is specifically designed for CPU architectures. 
By relying on regular array operations (including global mask construction and prefix scan), it produces compact and sequential memory layouts that align well with CPU characteristics (\textit{e.g.}, strong cache hierarchies). 
As a result, the scan-based implementation achieves efficient performance on CPUs.
Details of the parallel scan-based construction are given in Section~\ref{sec:scan_based_sparse_implementation}.

In contrast to the scan-based implementation, we develop a different strategy for GPU architectures. 
GPUs are efficient at executing many lightweight threads in parallel, but less efficient for global synchronization operations over the whole grid. 
These characteristics motivate a different approach for constructing the active grid. 
Instead of forming a global view of the grid, the hash-based implementation (Fig.~\ref{fig:mpm_sparse}c) constructs the active-node set in a fully local and parallel manner. 
For each particle, the grid nodes within its support are enumerated. 
A unique key is generated from the node's location (\textit{i.e.}, the integer index tuple $(i, j, k)$). 
Each key is then inserted into a hash table that maps node keys to compact indices. 
During insertion, if a key is encountered for the first time, the index is increased and the new value is assigned to that node. 
Instead, if the key already exists (\textit{e.g.}, the node is already checked), no new index is created. 
In this way, the hash table ensures the uniqueness of active nodes and performs on-the-fly index assignment. 
This approach avoids any global scan operation over the whole grid.
We refer to it as the hash-based approach since it relies on a hash table for indexing. 
This implementation is specifically designed for GPU architectures and is well aligned with the strengths of GPUs (massively parallel execution).
Details of the method and implementation are provided in Section~\ref{sec:hash_based_sparse_implementation}.

\subsection{Examples}
\begin{figure}[!ht]
    \centering
    \includegraphics[width=0.7\textwidth]{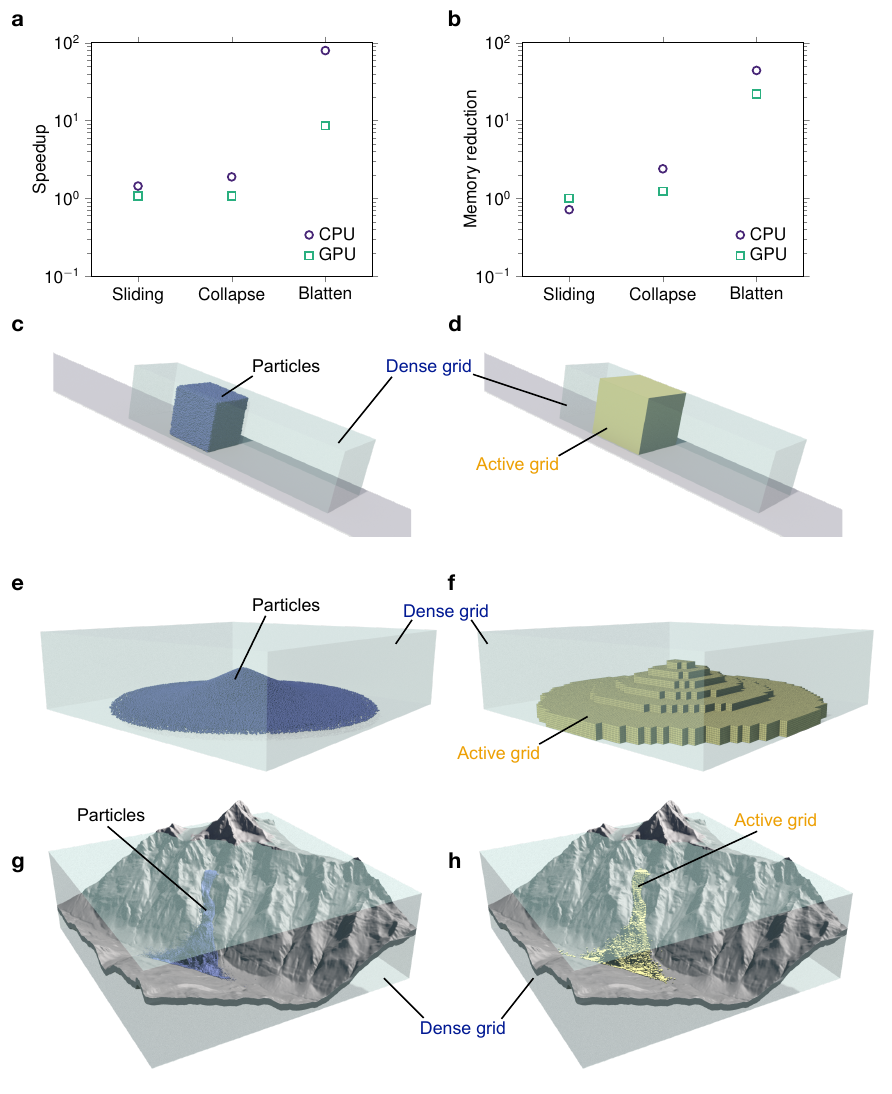}
    \caption{Performance and sparsity in three examples. a, b. Speedup (a) and memory reduction ratio (b) achieved by sparse MPM relative to the dense formulation for three examples: sliding box, granular collapse, and the Blatten landslide. Speedup is defined as the ratio of dense runtime to sparse runtime. Memory reduction is defined as the ratio of memory usage of the dense formulation to that of the sparse formulation. Open circles denote CPU results (scan-based approach), and open squares denote GPU results (hash-based approach). c, d. Sliding box example. c. Particle distribution within a dense background grid. d. Active grid region. e, f. Granular collapse example. e. Particle distribution over a larger dense background grid. f. Active grid region. g, h. Blatten landslide example. g. Particle distribution within a large dense background grid. h. Active grid region.}
    \label{fig:examples}
\end{figure}

We use three representative examples with increasing levels of sparsity to show the effectiveness of the sparse MPM. 
These examples include: a sliding box (Fig.~\ref{fig:examples}c, d), a granular collapse (Fig.~\ref{fig:examples}e, f), and a large-scale simulation of the 2025 Blatten landslide, a recent rock-ice avalanche in Switzerland (Fig.~\ref{fig:examples}g, h). 
They range from relatively dense configurations to highly sparse, large-scale geophysical flows.
To quantify the degree of sparsity, we define the sparsity ratio as
\begin{equation}
    r_{\mathrm{active}} = \min_t \dfrac{n_{\mathrm{dense}}}{n_{\mathrm{active}}(t)},
\end{equation}
where $n_{\mathrm{active}}(t)$ is the number of active grid nodes at time $t$, and $n_{\mathrm{dense}}$ is the number of grid nodes in the dense background grid. 
A larger sparsity ratio indicates stronger sparsity. 
The sparsity ratios increases from 2.9 for the sliding box to 5.5 for the granular collapse and 373 for the Blatten landslide.

Fig.~\ref{fig:examples}a and b summarize the key performance results. 
We show the speedup and memory reduction achieved by sparse MPM relative to the standard dense implementation (where the background grid is constructed as a bounding box, covering the potential domain that particles may move during the whole simulation steps). 
Here, speedup is defined as the ratio of dense runtime to sparse runtime, and memory reduction is defined as the ratio of dense memory usage to sparse memory usage.
CPU results, obtained using the scan-based implementation, are shown with open circles. 
GPU results, obtained using the hash-based implementation, are shown with open squares. 
The CPU implementation is based on our open-source framework \textit{Matter}~\citep{blatny2025matter,lars_blatny_2026_21757524} and available at \url{https://github.com/larsblatny/matter/}. The GPU implementation is built upon the open-source \textit{GeoWarp}~\citep{zhao2026geowarp,yidong_zhao_2026_21758667} and available at \url{https://github.com/Yidong-ZHAO/sparse_MPM}. 
All simulations are performed on a workstation equipped with an Intel Core Ultra 9 285K CPU (24 cores) and an NVIDIA RTX 5070 Ti GPU (17 GB memory). 
The CPU implementation is run using 8 threads.

For the sliding box example, the domain is only mildly sparse. 
As shown in Fig.~\ref{fig:examples}c, d, the particles occupy a relatively large portion of the whole computational domain. 
In this case, sparse MPM performs similarly to the dense baseline (with speedup and memory reduction ratio close to one).

For the granular collapse (Fig.~\ref{fig:examples}e, f), sparsity becomes stronger. 
The active region remains localized while the background grid must still cover a larger domain to accommodate potential spreading. 
As a result, sparse MPM provides measurable gains in both computational time and memory usage.

This effect becomes even more significant in the Blatten landslide (Fig.~\ref{fig:examples}g, h), where particles occupy only a small fraction of the large computational domain. Consequently, most grid nodes remain inactive. Sparse MPM substantially improves both computational efficiency and memory usage, with speedup and memory reduction ratios increasing by one to two orders of magnitude.
Interestingly, the CPU speedup (open circles in Fig.~\ref{fig:examples}a) is more significant than the GPU speedup.
This difference can be attributed to the different baseline costs of the dense MPM on each architecture. 
On the CPU, the dense implementation suffers more from computations over inactive regions because of limited thread-level parallelism and a stronger dependence on memory locality. 
Consequently, eliminating these unnecessary computations through the sparse approach results in larger relative performance gains. 
In contrast, the GPU partially mitigates the cost of inactive regions through massive parallelism, so the relative performance gain is smaller, although still significant.

\subsubsection{Low sparsity examples}
\begin{figure}[!ht]
    \centering
    \includegraphics[width=0.95\textwidth]{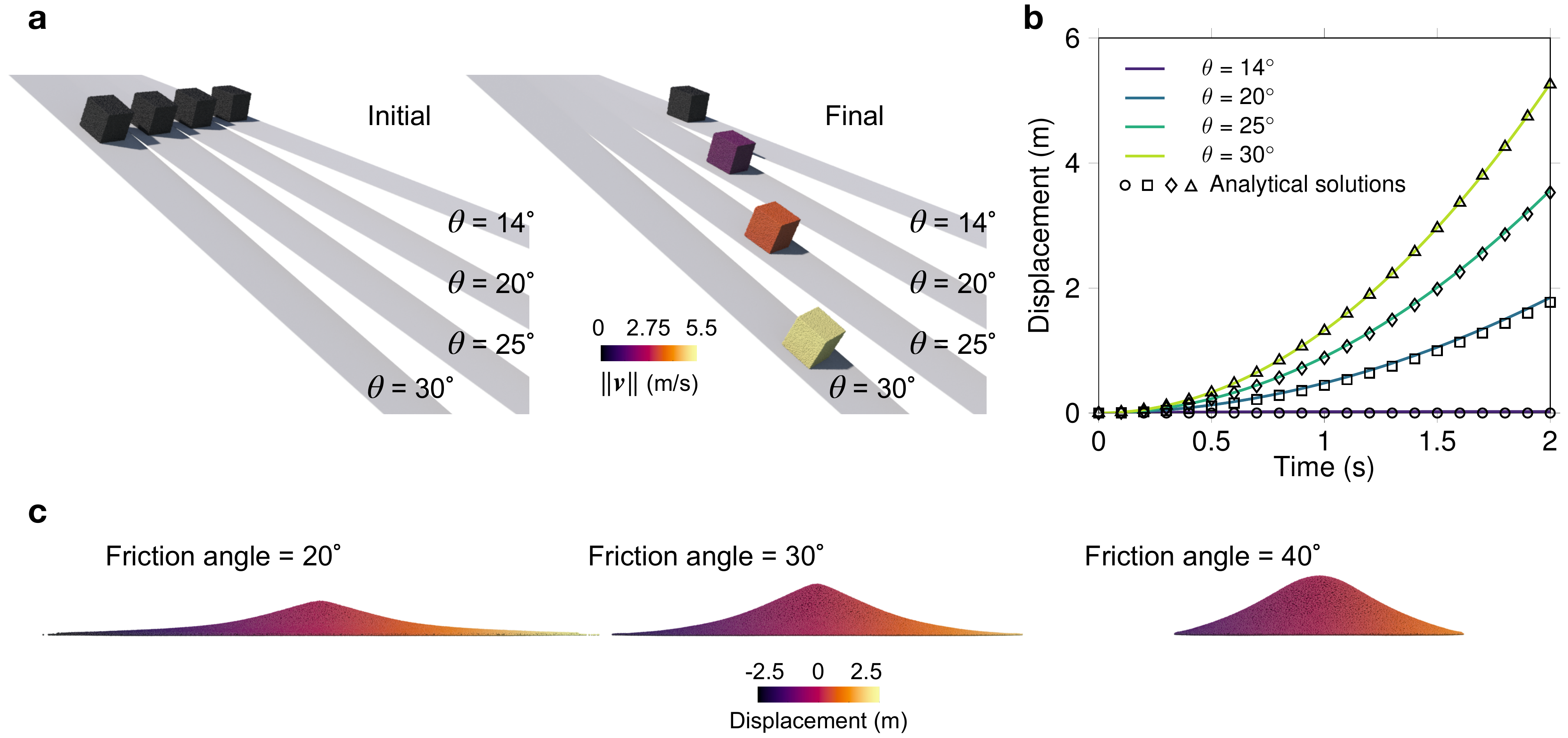}
    \caption{Examples with limited sparsity. a. Sliding box problems on inclined planes with different slope angles ($\theta = 14^\circ$, $20^\circ$, $25^\circ$, $30^\circ$). Left: initial configuration. Right: final configuration colored by velocity magnitude. b. Displacement-time curves for the sliding box problem. Solid lines denote simulation results, and markers denote analytical solutions. Great agreement is observed across all slope angles. c. Granular collapse under different internal friction angles ($20^\circ$, $30^\circ$, $40^\circ$). Colors indicate particle displacement. Lower friction angles result in longer runout distances and flatter deposits, while higher friction angles lead to more localized and steeper deposits.}
    \label{fig:examples_less_sparse}
\end{figure}

We first consider two benchmark problems with relatively low to moderate sparsity: a sliding box on an inclined plane and a granular column collapse (Fig.~\ref{fig:examples_less_sparse}). 
These examples aim to confirm that sparse MPM produces identical solutions as the standard dense implementation while introducing only a small overhead when sparsity is limited.

The sliding box example (Fig.~\ref{fig:examples_less_sparse}a) simulates blocks placed on inclined planes with different angles ($\theta = 14^\circ$, $20^\circ$, $25^\circ$, $30^\circ$). 
The block motion (displacement) is governed by Coulomb friction, and analytical solutions for displacement are available. 
The friction coefficient between the block and plane is $\mu = \tan(15^\circ) = 0.268$. 
In this case, the block should stick with $\theta = 14^\circ$ and slide for the remaining angles $\theta = 20^\circ$, $25^\circ$, $30^\circ$. 
Fig.~\ref{fig:examples_less_sparse}b compares the simulated displacements with the analytical solutions. 
The sparse and dense MPM results are identical, and both agree with the analytical solutions for all slope angles.  
In this example, the problem is only mildly sparse (with a sparsity ratio of 2.9). 
The particles occupy a relatively large portion of the grid during motion. 
As a result, the sparse framework yields performance similar to the dense formulation, as shown in Fig.~\ref{fig:examples}a, b.

The granular collapse example (Fig.~\ref{fig:examples_less_sparse}c) further tests the behavior under large deformation with moderate sparsity (with a sparsity ratio of 5.5). A granular column collapses under gravity for different internal friction angles. 
The final deposit profiles are shown in Fig.~\ref{fig:examples_less_sparse}c. 
As the friction angle increases, the final configuration becomes steeper and more localized.
As in the sliding box case, the sparse and dense MPM simulations produce identical results.
The difference is computational efficiency.
The active region remains localized, while the dense background domain must be large enough to accommodate potential spreading. Thus, sparse MPM demonstrates noticeable reductions in both computational time and memory usage (see Fig.~\ref{fig:examples}a, b).

Together, these two examples confirm that the sparse framework does not affect the solutions of standard MPM. 
When sparsity is limited, as in the sliding box, the additional overhead of constructing the sparse grid offsets most of the benefit.
As sparsity increases, the advantages of the sparse framework become progressively more significant. 
This will be demonstrated clearly in the next example.

\subsubsection{Blatten landslide simulation with strong sparsity}
\begin{figure}[!ht]
    \centering
    \includegraphics[width=1\textwidth]{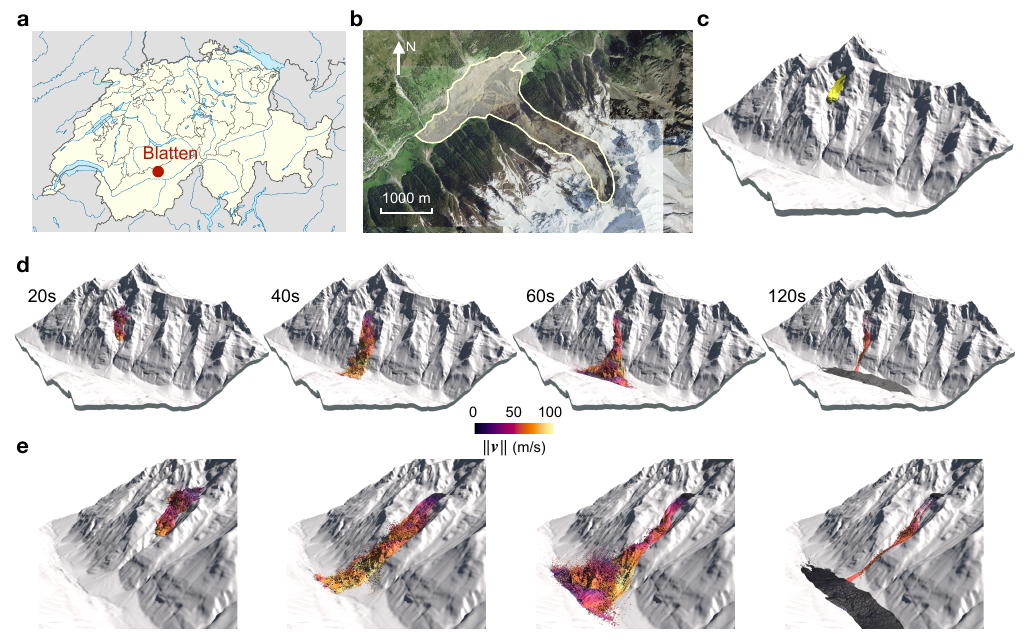}
    \caption{Large-scale simulation of the Blatten landslide. a. Location of the Blatten landslide in Switzerland (\url{https://commons.wikimedia.org/wiki/File:Switzerland\_location\_map.svg}). b. Satellite image of the study area with the landslide extent outlined (\url{https://map.geo.admin.ch/}). c. Three-dimensional terrain model with the initial release zone highlighted. d. The simulated progression of the landslide at $t = 20$, $40$, $60$, and $120$ s, showing propagation, spreading, and deposition processes. Colors indicate velocity magnitude. e. Zoomed-in views of the landslide progression corresponding to d.}
    \label{fig:examples_blatten}
\end{figure}

Next, we consider a real-world, large-scale geophysical flow with strong sparsity: the Blatten landslide (Fig.~\ref{fig:examples_blatten}). 
This example represents a challenging case in which the potential computational domain is large, while the flowing material occupies only a small fraction of the domain throughout the simulation. 

A rock-ice avalanche occurred at Birch Glacier in Switzerland on 28 May, 2025, where the geographical location is shown in Fig.~\ref{fig:examples_blatten}a. 
The landslide, with an estimated volume of $9.3 \times 10^6 \, \mathrm{m}^3$, destroyed much of the Blatten village~\citep{farinotti2025fact,kang2026frictional}. 
The avalanche moves down over an elevation difference of approximately 1000 m, and the final deposit is approximately 2 km long and 50 to 200 m wide~\citep{srf2025blatten}.
Fig.~\ref{fig:examples_blatten}b shows satellite imagery following the event. 
The three-dimensional terrain model and the initial release zone (highlighted in yellow) are shown in Fig.~\ref{fig:examples_blatten}c. 
The dense computational domain is chosen to cover the full potential runout area, as shown in Fig.~\ref{fig:examples}g. 
As a result, the domain is significantly larger than the region actually occupied by the flowing material, which leads to a highly sparse configuration.

The simulated motion of the landslide is shown in Fig.~\ref{fig:examples_blatten}d, e at different times. 
The flow initiates from the release zone, accelerates downslope, and progressively spreads into the valley. 
The color map indicates velocity magnitude, highlighting the rapid motion during the early stages (reaching approximately 100$\,$m/s) and the gradual deceleration as the material deposits. 
Notably, despite the large potential flow area and computational domain, the active region remains confined to a small fraction of the domain throughout the simulation. 
Such a small fraction indicates a strong sparsity of the problem (with a sparsity ratio of nearly 400).

\begin{figure}[!ht]
    \centering
    \includegraphics[width=0.9\textwidth]{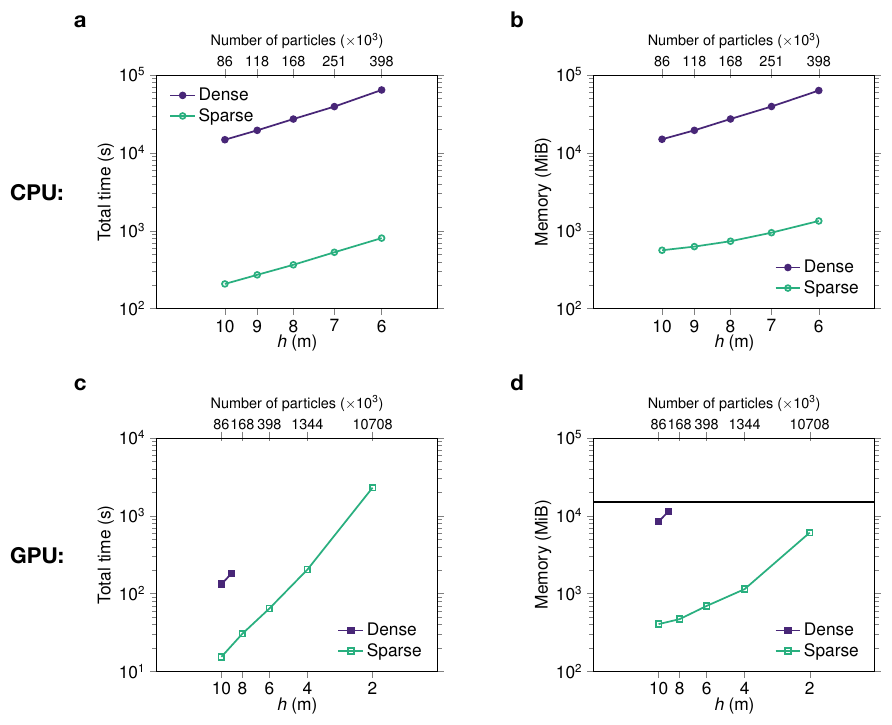}
    \caption{Scaling of computational cost and memory usage in the Blatten landslide example. a, b. Total computational time (a) and memory usage (b) as a function of background grid size $h$ (bottom row) and the corresponding number of particles (top row) for CPU simulation. Results are shown for both dense and sparse MPM. c, d. Total computational time (c) and memory usage (d) for GPU simulation. The horizontal line in d indicates the available GPU memory limit.}
    \label{fig:examples_blatten_scalability}
\end{figure}

The strong sparsity has a significant impact on computational performance. 
Fig.~\ref{fig:examples_blatten_scalability} shows the scaling of computational time and memory usage with grid resolution (indicated using the grid spacing $h$) for both dense and sparse schemes on CPU (Fig.~\ref{fig:examples_blatten_scalability}a, b) and GPU (Fig.~\ref{fig:examples_blatten_scalability}c, d). 
As shown in Fig.~\ref{fig:examples_blatten_scalability}a, c, the total computational time of sparse MPM is reduced by more than an order of magnitude compared to dense MPM. 
The time increases approximately linearly over the tested refinement range.
Fig.~\ref{fig:examples_blatten_scalability}b, d shows that memory usage is significantly reduced. Such a reduction in memory enables simulations that would otherwise exceed hardware limits. 
In particular, the dense GPU implementation quickly reaches the available memory limit (indicated by the horizontal line in Fig.~\ref{fig:examples_blatten_scalability}d), while the sparse formulation remains well within this memory constraint. 
We shall note that the simulation with the finest discretization (with $h = 2$ m) is possible on a single GPU only with sparse MPM.
As a practical reference, we also compare the sparse MPM on GPU with Houdini's MPM solver, a popular commercial software for MPM simulation~\citep{sidefx_houdini}.
Our implementation runs more than 1.5 times faster under the same setup.

The performance gains are more significant on the CPU than on the GPU, consistent with the observations in Fig.~\ref{fig:examples}a, b. 
On the CPU, the dense formulation incurs significant overhead due to operations and memory allocation over inactive regions. 
This makes it particularly sensitive to sparsity. 
Sparse MPM effectively eliminates these unnecessary memory allocations and computations, and leads to larger relative speedups (around two orders of magnitude in total runtime). 
On the GPU, although the dense formulation benefits from massive parallelism, sparse MPM still achieves significant improvements (on the order of one magnitude in total runtime) by reducing memory traffic and focusing computation on the active region.

Overall, the Blatten landslide example demonstrates the full potential of the proposed sparse framework for highly sparse problems. 
In these cases, the method provides significant reductions in computational cost and memory usage. 
Such reductions make high-resolution simulations possible on both CPU and GPU platforms.

\section{Methods}
In this section, we give a detailed introduction to the material point method formulation and the two implementations of the sparse MPM (scan-based and hash-based).

\subsection{Material point method formulation}
We adopt the material point method (MPM) for large-deformation continuum problems. 
The formulation presented here follows the standard updated Lagrangian scheme with explicit Euler time integration~\citep{nguyen2023material}.
Other variants, such as total Lagrangian and implicit schemes, are different in equations and solution strategies~\citep{guilkey2003implicit,de2020total}.
The proposed sparse framework is independent of these choices because it only changes how active grid nodes are stored and indexed.
Let $\Omega$ denote the current domain occupied by the material. 
The balance of linear momentum is written as
\begin{equation}
    \rho \dfrac{\mathrm{D} \boldsymbol{v}}{\mathrm{D} t} = \nabla \cdot \boldsymbol{\sigma} + \rho \boldsymbol{g},
\end{equation}
where $\rho$ is the density, $\boldsymbol{v}$ is the velocity, $\dfrac{\mathrm{D} \left(\cdot\right)}{\mathrm{D} t}$ indicates the material derivative, $\boldsymbol{\sigma}$ is the Cauchy stress tensor, and $\boldsymbol{g}$ is the body force per unit mass.

MPM solves this continuum problem using a hybrid Lagrangian-Eulerian discretization. 
The material is discretized into a set of Lagrangian particles. 
Each particle $p$ carries the material quantities, including its position $\boldsymbol{x}_p$, mass $m_p$, volume $V_p$, velocity $\boldsymbol{v}_p$, stress $\boldsymbol{\sigma}_p$, and deformation gradient $\boldsymbol{F}_p$. 
In contrast, the governing equations are solved on an auxiliary background grid. 
We denote a grid node by its integer grid tuple 
\begin{equation}
    \boldsymbol{n} = (i, \, j, \, k),
\end{equation}
where $i$, $j$, and $k$ are the integer indices of the node along the $x$, $y$, and $z$ directions, respectively. 
The corresponding physical position of the node is denoted by $\boldsymbol{x}_{\boldsymbol{n}} := (i\cdot h, \, j \cdot h, \, k \cdot h),$ where $h$ is the grid spacing (assumed equal in three directions). 
Communication between particles and grid nodes is achieved through interpolation functions $N_{\boldsymbol{n}}(\boldsymbol{x}_p)$ and their gradients $\nabla N_{\boldsymbol{n}}(\boldsymbol{x}_p)$. 
Different interpolation functions lead to different support sizes, that is, different sets of grid nodes over which the interpolation values are nonzero. 
In this work, we adopt quadratic B-splines~\citep{steffen2008analysis}. Other interpolation schemes, such as the generalized interpolation material point method (GIMP), convected particle domain interpolation (CPDI), and moving least squares MPM (MLS-MPM), can also be used~\citep{bardenhagen2004generalized,sadeghirad2011convected,sadeghirad2013second,hu2018moving}. The sparse framework is independent of the specific interpolation scheme.

Because the interpolation functions have compact support, each particle interacts with only a small number of neighboring grid nodes. 
For a given particle $p$, we denote this local interpolation support by 
\begin{equation}
    \mathcal{S}(p) = \{\boldsymbol{n} | N_{\boldsymbol{n}}(\boldsymbol{x}_p) \neq 0\}.
\end{equation}
This local support is key to the sparse formulation developed in this work. 
A grid node is said to be active if it belongs to the support of at least one particle. 
The set of active nodes at a given time step is therefore 
\begin{equation}
    \mathcal{A} := \bigcup_p \mathcal{S}(p).
\end{equation}
The sparse framework aims to allocate memory and perform grid-based computations only on this active-node set.
The size of this set, $|\mathcal{A}| = n_{\mathrm{active}}$, is the number of active nodes.

At each time step, MPM involves three steps: particle-to-grid transfer (P2G), grid update, and grid-to-particle transfer (G2P). 
In this work, particle-grid transfer applies the affine particle-in-cell (APIC) scheme~\citep{jiang2015affine} for reducing numerical dissipation and improving momentum conservation.
During P2G, particle quantities (\textit{e.g.}, mass and momentum) are transferred to grid nodes as 
\begin{align}
    &m_{\boldsymbol{n}} = \sum_p N_{\boldsymbol{n}}(\boldsymbol{x}_p) m_p, \\
    &m_{\boldsymbol{n}} \boldsymbol{v}_{\boldsymbol{n}} = \sum_p N_{\boldsymbol{n}}(\boldsymbol{x}_p) m_p \left[\boldsymbol{v}_p + \boldsymbol{C}_p (\boldsymbol{x}_{\boldsymbol{n}} - \boldsymbol{x}_p)\right],
\end{align}
where $\boldsymbol{C}_p$ denotes the affine velocity matrix in APIC.
Then the nodal velocities are obtained through the following normalization
\begin{equation}
    \boldsymbol{v}_{\boldsymbol{n}} = \dfrac{m_{\boldsymbol{n}} \boldsymbol{v}_{\boldsymbol{n}}}{m_{\boldsymbol{n}}}.
\end{equation}

The internal force at node $\boldsymbol{n}$ is computed as 
\begin{equation}
    \boldsymbol{f}_{\boldsymbol{n}}^{\mathrm{int}} = -\sum_p \nabla N_{\boldsymbol{n}}(\boldsymbol{x}_p) \boldsymbol{\sigma}_p V_p,
\end{equation}
and the external force is written as
\begin{equation}
    \boldsymbol{f}_{\boldsymbol{n}}^{\mathrm{ext}} = \sum_p N_{\boldsymbol{n}}(\boldsymbol{x}_p) m_p \boldsymbol{g}.
\end{equation}

Using an explicit Euler time integration scheme, the nodal momentum balance is written as:
\begin{equation}
    m_{\boldsymbol{n}} \dfrac{\boldsymbol{v}_{\boldsymbol{n}}^{n+1}-\boldsymbol{v}_{\boldsymbol{n}}}{\Delta t} = \boldsymbol{f}_{\boldsymbol{n}}^{\mathrm{int}} + \boldsymbol{f}_{\boldsymbol{n}}^{\mathrm{ext}},
\end{equation}
where $\Delta t$ denotes the time interval. 
Then the updated nodal velocity is obtained as
\begin{equation}
    \boldsymbol{v}_{\boldsymbol{n}}^{n+1} = \boldsymbol{v}_{\boldsymbol{n}} + \dfrac{\Delta t}{m_{\boldsymbol{n}}} \left( \boldsymbol{f}_{\boldsymbol{n}}^{\mathrm{int}} + \boldsymbol{f}_{\boldsymbol{n}}^{\mathrm{ext}} \right).
\end{equation}
The superscript $n+1$ denotes the next time step, at which the quantities are unknown (\textit{i.e.}, what we want to solve for). 
After updating the grid velocities, they are interpolated back to the particles through G2P using the same interpolation functions as in P2G. 
The particle velocity is updated as
\begin{equation}
    \boldsymbol{v}_p^{n+1} = \sum_{\boldsymbol{n} \in \mathcal{S}(p)} N_{\boldsymbol{n}}(\boldsymbol{x}_p) \boldsymbol{v}_{\boldsymbol{n}}^{n+1},
\end{equation}
and the particle position is updated by
\begin{equation}
    \boldsymbol{x}_p^{n+1} = \boldsymbol{x}_p + \Delta t \boldsymbol{v}_p^{n+1}.
\end{equation}
For the APIC scheme, the affine velocity matrix is updated through:
\begin{equation}
    \boldsymbol{C}_p^{n+1} = \left( \sum_{\boldsymbol{n} \in \mathcal{S}(p)} N_{\boldsymbol{n}}(\boldsymbol{x}_p) \boldsymbol{v}_{\boldsymbol{n}}^{n+1} \otimes (\boldsymbol{x}_{\boldsymbol{n}} - \boldsymbol{x}_p) \right) \boldsymbol{D}_p^{-1},
\end{equation}
where
\begin{equation}
    \boldsymbol{D}_p = \sum_{\boldsymbol{n} \in \mathcal{S}(p)} N_{\boldsymbol{n}}(\boldsymbol{x}_p) (\boldsymbol{x}_{\boldsymbol{n}} - \boldsymbol{x}_p) \otimes (\boldsymbol{x}_{\boldsymbol{n}} - \boldsymbol{x}_p).
\end{equation}
Lastly, the deformation gradient is updated by
\begin{equation}
    \boldsymbol{F}_p^{n+1} = \left( \boldsymbol{1} + \Delta t \sum_{\boldsymbol{n} \in \mathcal{S}(p)} \boldsymbol{v}_{\boldsymbol{n}}^{n+1} \otimes \nabla N_{\boldsymbol{n}}(\boldsymbol{x}_p) \right) \boldsymbol{F}_p,
\end{equation}
which is then used to update the particle stress based on the specific constitutive model. 
In this work, the sliding box example uses an elastic model, while the granular collapse and the landslide examples use a Drucker-Prager elastoplastic model~\citep{drucker1952soil}. 
Other constitutive models can be incorporated in the same framework. 
Readers are referred to~\citep{de2011computational,borja2013plasticity} for more details.

This formulation shows that only nodes in $\mathcal{A}$ participate in particle-grid transfer and grid update. 
Nevertheless, in the conventional dense formulation, all grid nodes in the potential computational domain are still allocated. 
The mismatch between the small active-node set and the much larger dense grid is the source of the unnecessary memory allocation and computational cost. 
In the sparse framework, storage and grid-based computations are instead restricted to active nodes only. 
To achieve this, the key task is to construct not only the active-node set $\mathcal{A}$, but also a compact indexing map
\begin{equation}
    \phi:\mathcal{A}\rightarrow \{0,1,\dots, |\mathcal{A}|-1\},
\end{equation}
which assigns each active node $\boldsymbol{n} \in \mathcal{A}$ a unique scalar index for sparse storage. 
Here, $\boldsymbol{n} = (i, \, j, \, k)$ identifies a node in the structured background grid, while $\phi(\boldsymbol{n})$ identifies the same node in compact sparse storage. 
The following subsections describe two implementations of this construction: a scan-based approach and a hash-based approach.

\subsection{Scan-based sparse implementation}
\label{sec:scan_based_sparse_implementation}
The scan-based implementation adopts a structured dense-to-sparse transformation. 
Its goal is to construct the active-node set $\mathcal{A}$ and the associated compact indexing map $\phi$ using regular array operations.

We begin by defining a binary activity mask over the candidate grid. 
For each grid node $\boldsymbol{n}$, the mask value $\chi_{\boldsymbol{n}}$ is written as
\begin{equation}
    \chi_{\boldsymbol{n}} =\begin{cases}1, & \boldsymbol{n} \in \mathcal A,\\0, & \text{otherwise}.\end{cases}
\end{equation}
The mask is constructed by looping over all particles and all nodes in their support. 
For each particle $p$, all nodes $\boldsymbol{n} \in \mathcal{S}(p)$ are visited, and the corresponding entries of the mask are marked as active (by setting $\chi_{\boldsymbol{n}}$ to 1). 
After all particles and their support nodes have been checked, the binary mask provides a dense representation of the active-node set.

Once the mask is constructed, a prefix scan is applied to obtain a compact indexing map. 
A prefix scan is an operation that replaces each entry of an array with the cumulative sum of all preceding entries. 
Specifically for this binary mask, the operation counts how many active nodes appear before a given grid node in the chosen traversal order. 
This cumulative count can therefore be the compact storage index for the active node.
Let $c_{\boldsymbol{n}}$ denote the prefix scan of $\chi_{\boldsymbol{n}}$. 
Then, for each active node $\boldsymbol{n} \in \mathcal{A}$, the compact index is assigned as
\begin{equation}
    \phi(\boldsymbol{n}) = c_{\boldsymbol{n}}.
\end{equation}
Inactive nodes, for which $\chi_{\boldsymbol{n}} = 0$, are ignored in the summation computation. 
In this way, the prefix scan assigns consecutive indices to all active nodes in the range
\begin{equation}
    \phi(\boldsymbol{n}) \in \{0,1,\dots, |\mathcal{A}|-1\}.
\end{equation}
Lastly, the compact indexing map $\phi$ is used to allocate sparse storage for nodal quantities. 
Instead of storing grid quantities over the whole dense background grid, we allocate memory only for the $|\mathcal{A}|$ active nodes. 
In the subsequent P2G, grid update, and G2P steps, a grid node $\boldsymbol{n} \in \mathcal{S}(p)$ is first mapped to its compact index $\phi(\boldsymbol{n})$, and all nodal quantities are retrieved and updated in this reduced index space. 
Therefore, the sparse representation does not change the original MPM formulation while avoiding storing data for inactive nodes.

In principle, one could construct a binary mask for every grid node, apply a prefix scan over all nodes, and allocate memory only for the active nodes. 
In practice, however, a purely node-wise scan can introduce considerable overhead. 
It requires a large mask array and a large scan operation.
To reduce this overhead, the scan-based implementation used in this work constructs the sparse grid at the block level~\citep{setaluri2014spgrid}. 
The candidate grid is partitioned into small regular blocks, each of which contains $B \times B \times B$ grid nodes. 
A block is considered active if it contains at least one grid node influenced by particles. 
Instead of scanning a mask over all grid nodes, we scan a much smaller mask over grid blocks. 
Once active blocks are identified and indexed, all nodes inside each active block are allocated consecutively in memory.

This block-level construction has two main advantages. 
First, it reduces the size of the activity mask and the cost of the prefix scan, since the scan is performed over blocks rather than individual nodes. 
Second, it improves memory locality because neighboring grid nodes within the same block are stored consecutively. 
The tradeoff is that some inactive nodes inside active blocks may also be allocated. 
However, for a moderate block size (here we set $B = 4$), this local overhead is typically outweighed by the reduced indexing cost and improved cache efficiency.

Algorithm~\ref{alg:block_scan_sparse} summarizes the parallel block-level scan-based construction. 
The procedure first marks active blocks from particle supports. 
Then the compact indices for these blocks are generated using a parallel exclusive scan. 
The scan is implemented in three stages: each thread computes a local scan over a consecutive segment of the block mask, the thread-local sums are prefix-summed to obtain offsets, and these offsets are added back to the local scan results. 
After the compact block map is constructed, all nodes inside active blocks are stored in consecutive memory locations and accessed through the resulting sparse node map.
The scan-based implementation is released as part of the open-source codebase \textit{Matter}~\citep{blatny2025matter}, with the repository available at \url{https://github.com/larsblatny/matter/}.
\begin{algorithm}[t]
\small
\caption{Parallel block-level scan-based sparse grid construction}
\label{alg:block_scan_sparse}
\begin{algorithmic}[1]
\REQUIRE Particle positions $\{\boldsymbol{x}_p\}_{p=1}^{n_p}$, supports $\mathcal{S}(p)$, block size $B$, number of CPU threads $n_{\mathrm{th}}$
\ENSURE Compact block map $\Phi$, compact node map $\phi$, sparse grid storage

\STATE Construct candidate block domain $[\boldsymbol{b}_{\min},\boldsymbol{b}_{\max}]$ from the bounding box of particle supports
\STATE Initialize block activity mask \(\chi_{\boldsymbol{b}}\gets 0\) for all candidate blocks
\STATE $\chi_{\boldsymbol{b}} \gets$ mark\_active\_blocks$(\{\boldsymbol{x}_p\},\mathcal{S},B)$ \hfill // See Algorithm~\ref{alg:mark_active_blocks}
\STATE $(s_q,n_{\mathrm{block}}^{\mathrm{active}}) \gets$ parallel\_scan$(\chi_{\boldsymbol{b}},n_{\mathrm{th}})$ \hfill // See Algorithm~\ref{alg:parallel_scan}
\STATE $\Phi \gets$ construct\_compact\_block\_map$(\chi_{q},s_q)$ \hfill // See Algorithm~\ref{alg:construct_compact_block_map}
\STATE Allocate sparse grid arrays of size $n_{\mathrm{block}}^{\mathrm{active}}B^3$
\STATE For any node $\boldsymbol{n}$, its index can be computed based on the block index and local node index $\ell(\boldsymbol{n})$ as  $\phi(\boldsymbol{n})=\Phi(\boldsymbol{b}(\boldsymbol{n}))B^3+\ell(\boldsymbol{n})$, where $\ell(\boldsymbol{n})=(i-i//B \times B)+B\big((j-j//B \times B)+B(k-k//B \times B)\big)$
\end{algorithmic}
\end{algorithm}

\begin{algorithm}[t]
\small
\caption{mark\_active\_blocks}
\label{alg:mark_active_blocks}
\begin{algorithmic}[1]
\FORALL{particles $p$ in parallel}
    \FORALL{nodes $\boldsymbol{n}\in\mathcal{S}(p)$}
        \STATE Compute the block coordinates $\boldsymbol{b}(\boldsymbol{n})\gets \boldsymbol{n}//B = (i//B, \, j//B, \, k//B)$
        \STATE Set $\chi_{\boldsymbol{b}(\boldsymbol{n})}\gets 1$
    \ENDFOR
\ENDFOR
\end{algorithmic}
\end{algorithm}

\begin{algorithm}[t]
\small
\caption{parallel\_scan}
\label{alg:parallel_scan}
\begin{algorithmic}[1]
\STATE Flatten \(\chi_{\boldsymbol{b}}\) into a one-dimensional array \(\chi_q\), \(q=0,\ldots,n_{\mathrm{block}}-1\), where $n_{\mathrm{block}}$ indicates the number of candidate blocks
\FORALL{threads $t=0,\ldots,n_{\mathrm{th}}-1$ in parallel}
    \STATE Assign range $[q_t^{\mathrm{start}},q_t^{\mathrm{end}})$ to thread $t$
    \STATE $r\gets 0$
    \FOR{$q=q_t^{\mathrm{start}}$ to $q_t^{\mathrm{end}}-1$}
        \STATE Assign the cumulative sum of active blocks $s_q\gets r$
        \STATE Increase the cumulative sum $r\gets r+\chi_q$
    \ENDFOR
    \STATE Store the thread-local sum $T_t\gets r$
\ENDFOR
\STATE Compute thread offsets by prefix summing the thread-local sums: $O_t\gets \sum_{\tau<t}T_\tau$
\FORALL{threads $t=0,\ldots,n_{\mathrm{th}}-1$ in parallel}
    \FOR{$q=q_t^{\mathrm{start}}$ to $q_t^{\mathrm{end}}-1$}
        \STATE Add the offsets to the cumulative sum  $s_q\gets s_q+O_t$
    \ENDFOR
\ENDFOR
\STATE $n_{\mathrm{block}}^{\mathrm{active}}\gets \sum_{t=0}^{n_{\mathrm{th}}-1}T_t$
\end{algorithmic}
\end{algorithm}

\begin{algorithm}[t]
\small
\caption{construct\_compact\_block\_map}
\label{alg:construct_compact_block_map}
\begin{algorithmic}[1]
\STATE Allocate compact block list of size $n_{\mathrm{block}}^{\mathrm{active}}$
\FORALL{candidate block indices $q$ in parallel}
    \STATE Decode $q$ to block coordinates $\boldsymbol{b}$
    \IF{$\chi_q=1$}
        \STATE Assign compact block index $\Phi(\boldsymbol{b})\gets s_q$
        \STATE Store $\boldsymbol{b}$ in compact block list at position $s_q$
    \ELSE
        \STATE $\Phi(\boldsymbol{b})\gets -1$
    \ENDIF
\ENDFOR
\end{algorithmic}
\end{algorithm}

\subsection{Hash-based sparse implementation}
\label{sec:hash_based_sparse_implementation}
The hash-based implementation realizes the same sparse framework through a dynamic construction of the active-node set and its compact indexing map. 
In contrast to the scan-based implementation, which starts from a dense mask over the whole candidate grid, the hash-based implementation avoids any global grid view. 
Instead, it constructs $\mathcal{A}$ and the associated compact indexing map $\phi$ using on-the-fly insertion and lookup operations.

For each particle $p$, all grid nodes $\boldsymbol{n} \in \mathcal{S}(p)$ within its support are enumerated. 
Each node is identified by its integer grid tuple $\boldsymbol{n} = (i, \, j, \, k)$. 
Since this three-component integer identifier is inconvenient for direct storage, comparison, and lookup in a hash table, it is first converted into a unique scalar integer key. 
In our implementation, each integer component is shifted by a fixed bias $b$ and packed into a 64-bit integer as follows~\citep{setaluri2014spgrid}:
\begin{equation}
    \mathrm{Key}(\boldsymbol{n})=((i+b) \ll 2m)\;|\;((j+b) \ll m)\;|\;(k+b),
    \label{eq:key}
\end{equation}
where $m$ denotes the number of bits allocated to each integer, $\ll $ denotes bit shifting, and $|$ denotes the bitwise OR operation. 
The bias $b$ is introduced to handle negative grid indices.
This packing gives a unique scalar representation of each node, as long as the shifted coordinate lies in the range $0 \leq i+b,\, j+b,\, k+b < 2^m$, and the total number of packed bits satisfies $3m \leq 64$.
In this work, we use $m = 21$ and $b = 2^{20}$, which supports grid indices in the range $[-2^{20},\,2^{20}-1]$ along each coordinate direction.
Also, the bitwise operations used in the packing are extremely efficient and straightforward to implement.

Once a packed key represents a node, the next task is to determine whether this node has already been encountered. 
If not, we assign a new compact index to it. 
To this end, we use a hash table that stores the mapping from packed node keys to compact sparse indices. 
In this way, the hash table has two purposes: it detects whether an active node is new, and it constructs the indexing map $\phi$ on the fly. 
Although the packed key uniquely identifies a node, it is generally not suitable to use the key itself as a hash-table slot index.
The key space can be very large because it encodes three grid coordinates into a 64-bit integer.
Directly using these keys as array indices may require an impractically large table. 
Instead, a hash function is used to map the large key space into a much smaller array of table slots. 
The quality of this mapping is important for performance.
Nearby grid nodes often produce packed keys with similar bit patterns.
If these structured key values are mapped poorly to table slots, they can cluster and cause frequent collisions.
To reduce this effect, we apply a 64-bit integer mixing function~\citep{steele2014fast,stafford2011bitmixing}. 
The mixing function redistributes the bits of the packed key. 
Conceptually, this mixing function scrambles the structured key values so that even nearby nodes are mapped to well-distributed hash values. 
The hash value is written as:
\begin{equation}
    h(\boldsymbol{n})=\mathrm{Hash}(\mathrm{Key}(\boldsymbol{n})).
\end{equation}
The initial slot, $s(\boldsymbol{n})$, in the hash table is then obtained from the hash value through a bit mask:
\begin{equation}
    s(\boldsymbol{n})=h(\boldsymbol{n})\;\&\;(H-1),
\end{equation}
where $H$ is the capacity of the hash table (\textit{i.e.}, the total length of the hash table), and $\&$ denotes the bitwise AND operation.  
If the slot of the hash table is empty, the packed key is inserted, and a new compact index is assigned to the node. 
If the slot already contains the same key, this node has already been checked and it is skipped. 
Otherwise (the slot contains a different key), a collision is detected and the next slot is examined. 
In this work, collisions are resolved by open addressing with linear probing as:
\begin{equation}
    s \leftarrow (s+1)\;\&\;(H-1),
\end{equation}
which is repeated until either an empty slot is reached or the same key is found.

After all particles and associated nodes are checked, the compact indexing map $\phi$ is also constructed. 
Then the sparse storage is allocated only for the active nodes. 
The hash-based algorithm involves three steps: (1) enumerate support nodes from particles, (2) insert and construct compact indices using the hash table, and (3) allocate sparse arrays based on the final number of unique active nodes. 
The hash-based implementation uses the same block-level grid representation as the scan-based implementation. 
That is, the sparse grid is constructed from active blocks, and all nodes inside an active block are stored in consecutive memory locations. 

The implementation is summarized in Algorithm~\ref{alg:block_hash_sparse}. 
We first initialize an empty hash table and an active-block counter. 
Then, in parallel, particles enumerate their support nodes, map them to block keys (\textit{i.e.}, $\mathrm{Key}(\lfloor i/B \rfloor, \, \lfloor j/B\rfloor, \, \lfloor k/B\rfloor)$ using Eq.~\ref{eq:key}, where $\lfloor \cdot \rfloor$ indicates the floor operation), and insert these keys into the hash table. 
Each successful insertion creates one compact block index. 
We shall note that parallel insertion requires atomic operations because multiple GPU threads may attempt to insert the same active block simultaneously. 
The implementation uses atomic compare-and-swap to claim empty hash-table slots. 
Only the thread that successfully inserts a new key increments the active-block counter and assigns a new compact block index. 
After all particles have been checked, the active-block counter gives the number of active blocks. 
Then the sparse arrays are allocated with the same block-level layout described in the scan-based implementation.
The hash-based implementation is released as open-source code built on \textit{GeoWarp}~\citep{zhao2026geowarp} and is available at \url{https://github.com/Yidong-ZHAO/sparse_MPM}.
\begin{algorithm}[t]
\small
\caption{Parallel block-level hash-based sparse grid construction}
\label{alg:block_hash_sparse}
\begin{algorithmic}[1]
\REQUIRE Particle positions $\{\boldsymbol{x}_p\}_{p=1}^{n_p}$, particle supports $\mathcal{S}(p)$, block size $B$, initial hash-table capacity $H$
\ENSURE Compact block map $\Phi$, compact node map $\phi$, sparse grid storage

\STATE Initialize hash keys with the empty key
\STATE Initialize hash values with invalid block indices
\STATE Initialize active block list
\STATE Set $n_{\mathrm{block}}^{\mathrm{active}}\gets 0$
\STATE Set overflow flag to false
\STATE insert\_active\_blocks$(\{\boldsymbol{x}_p\},\mathcal{S},B,H)$ \hfill // See Algorithm~\ref{alg:insert_active_blocks_hash}
\IF{overflow flag is true}
    \STATE $(H,\Phi)\gets \mathrm{rebuild\_hash\_table}(H)$ \hfill // See Algorithm~\ref{alg:rebuild_hash_table}
\ENDIF
\STATE Allocate sparse grid arrays of size $n_{\mathrm{block}}^{\mathrm{active}}B^3$
\STATE For any node $\boldsymbol{n}$, its index can be computed based on the block index and local node index $\ell(\boldsymbol{n})$ as  $\phi(\boldsymbol{n})=\Phi(\boldsymbol{b}(\boldsymbol{n}))B^3+\ell(\boldsymbol{n})$, where $\ell(\boldsymbol{n})=(i-i//B \times B)+B\big((j-j//B \times B)+B(k-k//B \times B)\big)$
\end{algorithmic}
\end{algorithm}

\begin{algorithm}[t]
\small
\caption{insert\_active\_blocks}
\label{alg:insert_active_blocks_hash}
\begin{algorithmic}[1]
\FORALL{particles $p$ in parallel}
    \FORALL{nodes $\boldsymbol{n}\in\mathcal{S}(p)$}
        \STATE Compute the block coordinates $\boldsymbol{b}\gets\boldsymbol{n}//B = (i//B, \, j//B, \, k//B)$
        \STATE Pack block coordinates into a scalar key $K(\boldsymbol{b})$
        \STATE Compute hash value $h(\boldsymbol{b}) \gets \mathrm{Hash}(K(\boldsymbol{b}))$
        \STATE Compute initial slot $s\gets h(\boldsymbol{b})\ \&\ (H-1)$

        \FOR{$r=0$ to $n_{\mathrm{probe}}-1$}
            \STATE // $n_{\mathrm{probe}}$ indicates the maximum proving number
            \STATE $K_{\mathrm{old}}\gets
            \mathrm{atomic\_compare\_and\_swap} (\mathrm{hash\_keys}[s],\mathrm{EMPTY},K(\boldsymbol{b}))$

            \IF{$K_{\mathrm{old}}=\mathrm{EMPTY}$}
                \STATE $\Phi(\boldsymbol{b}) \gets \mathrm{atomic\_add}(n_{\mathrm{block}}^{\mathrm{active}},1)$
                \IF{$\Phi(\boldsymbol{b})$ is within the allocated block capacity}
                    \STATE $\mathrm{hash\_vals}[s]\gets \Phi(\boldsymbol{b})$
                    \STATE $\mathrm{active\_block\_keys}[\Phi(\boldsymbol{b})]\gets K(\boldsymbol{b})$
                \ELSE
                    \STATE Set overflow flag
                \ENDIF
                \STATE $\mathrm{inserted}\gets\mathrm{true}$
                \STATE \textbf{break}

            \ELSIF{$K_{\mathrm{old}}=K(\boldsymbol{b})$}
                \STATE The block has already been inserted
                \STATE $\mathrm{inserted}\gets\mathrm{true}$
                \STATE \textbf{break}

            \ELSE
                \STATE Resolve collision by linear probing: $s\gets (s+1)\ \&\ (H-1)$
            \ENDIF
        \ENDFOR

        \IF{$\mathrm{inserted}=\mathrm{false}$}
            \STATE Set overflow flag
        \ENDIF
    \ENDFOR
\ENDFOR
\end{algorithmic}
\end{algorithm}

\begin{algorithm}[t]
\small
\caption{rebuild\_hash\_table}
\label{alg:rebuild_hash_table}
\begin{algorithmic}[1]
\REPEAT
    \STATE Increase the hash-table capacity, e.g. $H\gets 2H$
    \STATE Reallocate hash keys and values with capacity $H$
    \STATE Reinitialize hash keys, hash values, active block list, and counter
    \STATE Set overflow flag to false
    \STATE insert\_active\_blocks$(\{\boldsymbol{x}_p\},\mathcal{S},B,H)$ \hfill // See Algorithm~\ref{alg:insert_active_blocks_hash}
\UNTIL{overflow flag is false}
\end{algorithmic}
\end{algorithm}

\subsection{Comparison of scan-based and hash-based implementations}
\label{sec:comparison}
\begin{figure}[ht!]
    \centering
    \includegraphics[width=0.9\textwidth]{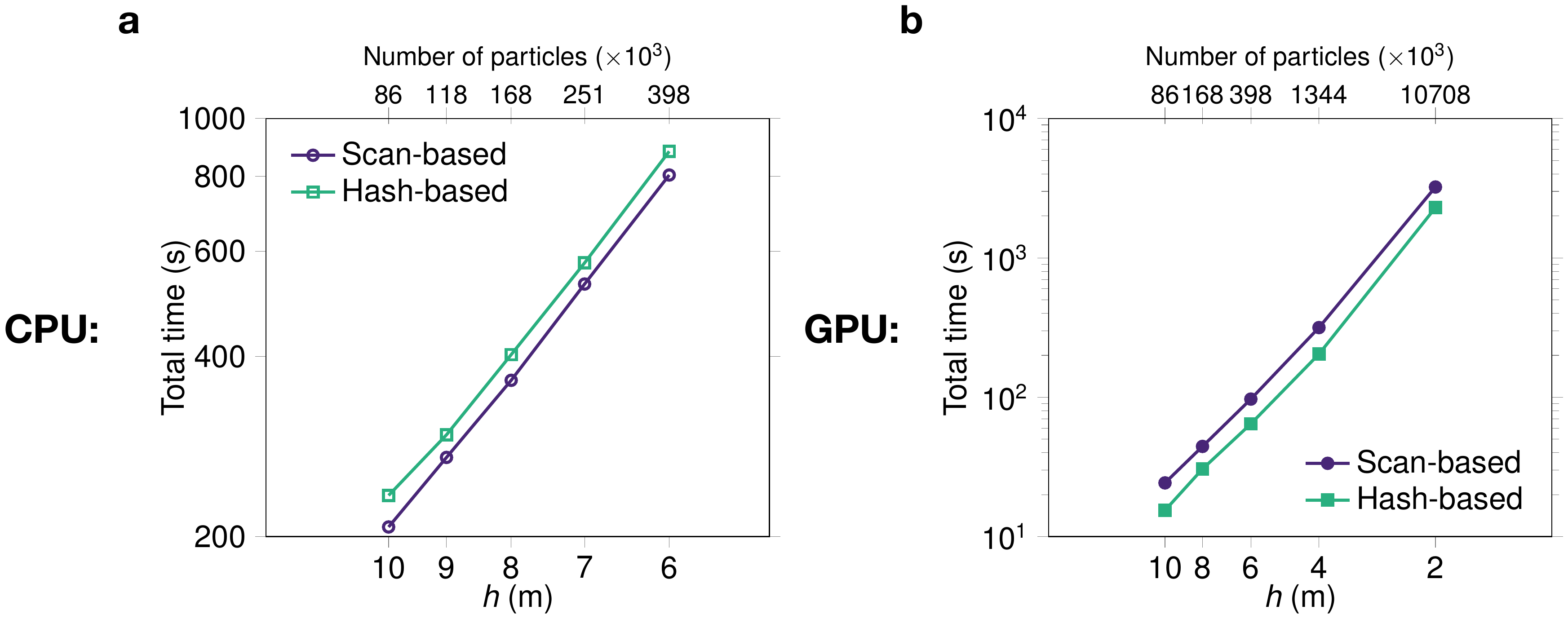}
    \caption{Comparison of scan-based and hash-based sparse implementations. $\mathbf{a}$. Total computational time of the scan-based and hash-based implementations on CPU for the Blatten landslide example. $\mathbf{b}$. Total computational time of the two implementations on GPU.}
    \label{fig:time_comparison}
\end{figure}

We further compare the scan-based and hash-based implementations on both CPU and GPU using the Blatten landslide example. 
The purpose of this comparison is to show how the same sparse framework performs when realized through different implementations. 
Fig.~\ref{fig:time_comparison}a, b show the total computational time on CPU and GPU, respectively. 
On the CPU, the scan-based implementation is consistently faster than the hash-based implementation. 
On the GPU, the opposite trend is observed. The hash-based implementation outperforms the scan-based implementation. 
Although this difference is less visually significant on the logarithmic scale in Fig.~\ref{fig:time_comparison}, the scan-based implementation reduces runtime by approximately 10\% on the CPU (Fig.~\ref{fig:time_comparison}a), while the hash-based implementation reduces runtime by approximately 35\% on the GPU (Fig.~\ref{fig:time_comparison}b). 
These observations are discussed in detail in Section~\ref{sec:sparse_mpm}. 
Overall, this comparison supports the use of architecture-specific implementations within the unified sparse framework.

\conclusions  
We have introduced a unified sparse background-grid framework for the material point method. 
The key idea is to restrict memory allocation and computation to grid nodes that are actively influenced by particles. 
Two architecture-specific implementations are developed: a scan-based implementation for CPUs and a hash-based implementation for GPUs. 
Through three examples with different sparsity levels, we show that this framework achieves substantial reductions in computational cost and memory usage in highly sparse scenarios while maintaining performance similar to the typical dense formulation in cases with low sparsity.

A key insight of this work is that the sparse framework can be separated from its implementation. 
The framework of an active-node set and its compact indexing is decoupled from the underlying data structures, unlike existing works such as~\citep{gao2018gpu,wang2020massively,qiu2023sparse,chen2025sparse}. 
Different sparse approaches have been proposed, each tailored to specific hardware architectures. The scan-based implementation adopts a global view with a binary mask. 
The compact indexing is then constructed through a prefix scan of the mask. 
It requires structured array operations and is well-suited to CPU architectures. 
In contrast, the hash-based implementation avoids a global scan and constructs the active-node set in a fully parallel manner. 
A hash table data structure is used to handle parallel on-the-fly index assignment, which aligns well with the strengths of GPU architectures. 
As a result, the relative efficiency of the two approaches depends on the hardware.
A quantitative comparison between them on CPU and GPU is provided. 
The separation between the sparse framework and its implementations also makes it easy to integrate into any existing MPM codes on different hardware platforms. 
We release the two implementations as open-source to provide practical references for incorporating the framework into MPM codes on different hardware platforms.

Future work may explore several directions. 
The current work focuses on single-node CPU and GPU architectures. 
Extending the framework to distributed-memory systems, such as multi-GPU or cluster environments~\citep{germain2000uintah,wang2020massively,qiu2023sparse}, would be a promising direction. 
It may require careful consideration of data partitioning and communication. 
The framework could also be applied to more general MPM formulations, such as multiphase coupling~\citep{yu2024semi,juel2026stabilized}.
These extensions can broaden the applicability of sparse MPM to a wide range of problems in geoscientific modeling.
More broadly, the proposed framework is not limited to MPM and may be applicable to a broader class of particle-grid or hybrid methods, such as the immersed boundary method~\citep{peskin2002immersed} and vortex-in-cell method~\citep{christiansen1973numerical}. 
Any method in which interactions are localized to a subset of grid nodes can potentially benefit from a similar sparse framework. 
The present work therefore contributes to sparse computation, where effort is focused only on regions that are physically active.

\codeavailability{The CPU implementation is available at \url{https://github.com/larsblatny/matter/}~\citep{lars_blatny_2026_21757524} (DOI:10.5281/zenodo.21757524) under the GNU General Public License v3.0 (GPL-3.0). The GPU implementation is available at \url{https://github.com/Yidong-ZHAO/sparse_MPM}~\citep{yidong_zhao_2026_21758667} (DOI:10.5281/zenodo.21758667) under the MIT License.} 













\authorcontribution{Y.Z. and J.G. conceptualized the study. Y.Z., C.J., and J.G. developed the methodology. Y.Z., L.B., and X.F. developed the software. Y.Z., L.B., M.M.J. wrote the original manuscript, and all authors revised the manuscript. C.J. and J.G. supervised the study.} 

\competinginterests{The authors declare no competing interests.} 


\begin{acknowledgements}
Y.Z. gratefully acknowledges financial support from the Swiss National Science Foundation (grant number: CRSK-2\_237763).
L.B. gratefully acknowledges financial support from the Swiss National Science Foundation~(SNSF) through grant number~P500PT\_230265.
The authors used ChatGPT to assist with the writing during the preparation of the work. The authors reviewed and edited the content as needed and take full responsibility for the content of the article.
\end{acknowledgements}







\bibliographystyle{copernicus}
\bibliography{references.bib}

\end{document}